\documentstyle[prl,aps,epsfig]{revtex}
\twocolumn

\begin{document}

\title{ The Anderson transition: time reversal symmetry and universality}

\author{Keith Slevin}
\address{The Institute of Physical and Chemical Research, Hirosawa 2-1,
Wako-shi, Saitama 351-01, Japan}

\author{Tomi Ohtsuki}

\address{Department of Physics, Sophia University,
Kioi-cho 7-1, Chiyoda-ku, Tokyo 102, Japan}
\maketitle

\begin{abstract}
We report a finite size scaling study of the Anderson transition.
Different scaling functions and different
values for the critical exponent have been found,
consistent with the existence of the orthogonal
and unitary universality classes which occur in the
field theory description of the transition.
The critical conductance distribution at the
Anderson transition has also been investigated
and different distributions
for the orthogonal and unitary classes obtained.
\end{abstract}


\noindent
It is now widely accepted that the metal- insulator transition
for non- interacting electrons, the so called Anderson
transition \cite{KM}, is a continuous phase transition
in which static disorder plays a role
analogous to temperature in thermal phase transitions.
The field theoretical formulation of the problem \cite{EFETOV},
though not making reliable predictions for the
critical exponent \cite{WEGNER,HIKAMI},
does indicate that it should be possible to describe the
critical behavior within
a framework of three universality classes: orthogonal,
unitary and symplectic.
However, recent work put this idea in
question.
It was found that the scaling functions
for systems with orthogonal and unitary symmetry
could not be distinguished at an accuracy of
a few percent \cite{HKO}.
Nor could the values of the critical exponent
for the three universality classes be reliably
distinguished \cite{HKO,MAC,OOK,CD,TOHRU}.
This, together with recent work on the statistics of energy levels
in the vicinity of transition,
has prompted the suggestion \cite{HOFSTETTER} that the universality classes
predicted by the field theory do not correctly describe
the Anderson transition.

The two important symmetries in the field theory are
time reversal symmetry (TRS) and spin rotation symmetry (SRS).
The system is said to be in the orthogonal universality class if it
has both SRS and TRS,
in the unitary class if TRS is broken
and in the symplectic class if the system has TRS but SRS is broken.
The relevant terms in the Hamiltonian
are a coupling to an applied
magnetic field, which breaks TRS, and the spin orbit interaction,
which breaks SRS.

Here we focus on the breaking of TRS by a constant
applied magnetic field.
We report the results of Monte Carlo
studies which indicate that
the critical behavior, at least as far as
orthogonal and unitary symmetries are
concerned, is in
accord with the conventional universality classes.
By carrying out a precise study of the finite size scaling of
the electron localization length,
which is analogous to the correlation length in thermal transitions,
we have clearly differentiated the
scaling functions for the orthogonal and unitary
universality classes.
We have also estimated the critical exponents and
calculated confidence intervals for these estimates.
We find a statistically significant difference of about
$12\%$ between the values of the critical exponent in the
orthogonal and unitary classes.

To reinforce the above conclusion we have also simulated
the critical conductance
distribution of a disordered mesoscopic conductor.
Following the discovery of universal conductance fluctuations
it was realized that the conductance of a phase coherent system
is not self- averaging.
Extrapolating from the metallic
regime, it seems that the conductance fluctuations
at the critical point should be of the same order
of magnitude as the mean conductance and that the full
conductance distribution 
should be a more useful characteristic of the
critical point.
As we approach the critical point the localization length diverges
and the system becomes effectively self similar, we then expect
the conductance distribution to become independent of
system size and depend only on the universality class.
This expectation was borne out in our study where
we found different critical conductance distributions
depending on whether or not TRS is broken.

The model Hamiltonian used in this study describes
non- interacting electrons on a simple cubic lattice.
With nearest neighbor interactions only we have
\begin{equation}
\begin{array}{lll}
<\vec r| H | \vec r> & =  & V(\vec r) , \\
<\vec r| H | \vec r - \hat x> & =    & 1 , \\
<\vec r| H | \vec r - \hat y> & =    & 1 , \\
<\vec r| H | \vec r - \hat z> & =    & \exp(-i 2 \pi \phi x) ,
\end{array}
\label{model}
\end{equation}
where $\hat x$, $\hat y$ and $\hat z$ are the basis vectors of the
lattice.
The electrons are subject to
an external magnetic field applied in the $\hat y$ direction
whose strength is parameterized by the flux
$\phi$,
measured in units of the flux quantum
$h/e$,
threading a lattice cell.
The on site energies of the electrons $\{ V(\vec r) \}$
are assumed to be independently and identically distributed
with probability $p(V)\text{d}V $ where
\[
\begin{array}{llll}
p(V) & = &1/W & |V| \leq W/2  ,\\
                            & = & 0 & {\rm otherwise} .
\end{array}
\]

The critical point, scaling function and the value of the critical
exponent for several values of $\phi$ and Fermi energy $E_f$ 
(see Table \ref{T1}) are
determined
by examining the finite size scaling
\cite{GOLD} of
the localization length $\lambda$ for electrons on a
quasi-$1d$ dimensional bar of cross section $L \times L$.
The localization length $\lambda\equiv\lambda(E_f,\phi,W,L)$,
defined by
\begin{equation}
\lambda^{-1} = \lim_{L_z \rightarrow \infty} \frac{<-\ln  g(L_z)>}{2 L_z} ,
\label{lambda}
\end{equation}
where $L_z$ is the length of the bar and $g(L_z)$ is the
conductance of the bar measured in units of $(e^2/h)$,
can be evaluated by rewriting the
Schroedinger equation
as a product of transfer matrices; $\lambda^{-1}$
can then be determined to within
a specified accuracy using a standard technique \cite{MK}.
The accuracy used here ranges between
$0.1\%$ and $0.2\%$.

On intuitive grounds it has been argued \cite{Imry} that 
we should observe orthogonal scaling when $L\ll L_H$ and
unitary scaling when $L\gg L_H$ where $L_H$ is the
magnetic length.
For the lattice model (\ref{model}) $L_H=\sqrt{1/2\pi \phi}$.
For the smallest system size $L=6$ used here this criterion
yields a crossover flux $\phi_c\simeq 1/226$.
We thus expect to see clear unitary scaling behavior for 
cases Ua and Ub listed in Table \ref{T1}.

\begin{figure}
\epsfig{file=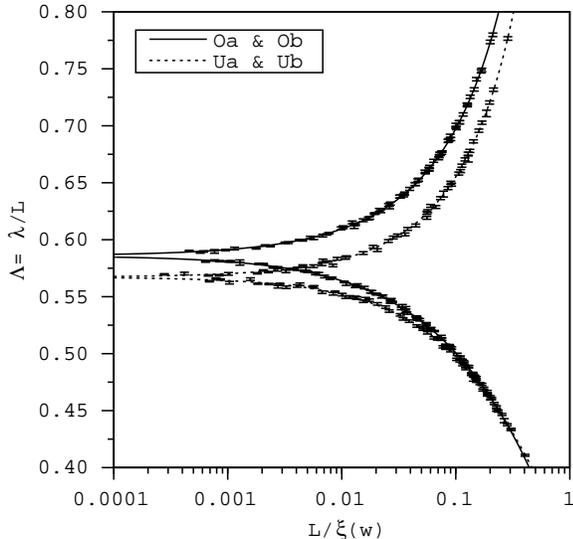,width=3.0in}
\caption{The scaling of the data for the parameter sets listed in Table
\ref{T1}.
The lines are the scaling functions given by (\ref{scalf+})
and (\ref{scalf-}).
Different scaling functions for the orthogonal and unitary universality
classes can be clearly distinguished.}
\label{F1}
\end{figure}

When the dimensionless quantity $\Lambda=\lambda/L$
is plotted against disorder $W$ for different cross sections $L$
the curves are found to have a common point of intersection;
this indicates the occurrence of the
metal- insulator
transition at a critical disorder $W_c$
in the $3d$ system which would be obtained by
letting $L \rightarrow \infty$.
Detailed analysis of the data is based on the following assumptions:
first that for $L$ finite $\Lambda$
is a smooth function of $W$ and $L$, second that the data obey a
one parameter scaling law
\begin{equation}
\Lambda(W,L) = f_{\pm}(L/\xi(w)) ,
\label{fssxi}
\end{equation}
where $w=(W_c-W)/W_c$ and the subscript $\pm$ refers to
$w>0$ and $w<0$,
and third that the length $\xi$ appearing in the scaling
law has a power law divergence close to $W_c$ of the form
$\xi(w) = \xi^{\pm} |w|^{-\nu}$.
This relation defines the critical exponent $\nu$ and introduces
two arbitrary constants $\xi^+$ $(w>0)$ and
$\xi^-$ $(w<0)$.
According to the Wegner scaling law \cite{WEGNER2} $\nu$
is related to the critical exponent
$s$ associated with the conductivity by $s=(d-2)\nu$ so that
$\nu=s$ in $d=3$.
The above assumptions imply that it should be possible to fit the data to
\begin{equation}
\Lambda_{\text{fit}}(W,L) = \Lambda_c + \sum_{n=1} A_n L^{n/\nu} w^n .
\label{fit}
\end{equation}
In practice we have truncated this series at $n=3$.
The relation between (\ref{fssxi}) and (\ref{fit}) can be
made apparent by writing
\begin{equation}
\begin{array}{ll}
f_{+} = \Lambda_c + \sum_{n=1} a_n \left( L/\xi \right)^{n/\nu},
&
a_n=A_n \left( \xi^+ \right)^{n/\nu}
\end{array} ,
\label{scalf+}
\end{equation}
\begin{equation}
\begin{array}{ll}
f_{-} = \Lambda_c + \sum_{n=1} b_n \left( L/\xi \right)^{n/\nu},
&
b_n=(-1)^n A_n \left( \xi^- \right)^{n/\nu}
\end{array} .
\label{scalf-}
\end{equation}
In principle $\xi^+$ and $\xi^-$ should depend on energy and
flux though 
the ``amplitude ratio'' $\xi^+/\xi^-$ may be universal
\cite{ZINN} so that $\xi^+$ and $\xi^-$
may not be independent.
Their absolute values cannot be determined using
the present method.
No relative variation
as a function of energy and
flux,
which would be apparent in the simulation, was detected.
Therefore for convenience we  set
$\xi^+=\xi^-=1$.
The most likely fit is determined by
minimizing the $\chi^2$-statistic
\begin{equation}
\chi^2 = \sum_m \left(
\frac{\Lambda_m-\Lambda_{\text{fit}}(W_m,L_m)}{\sigma_m} \right)^2 ,
\label{chi2}
\end{equation}
where the summation $m$ is over all data points
and $\sigma_m$ is the
error (standard deviation)
in the determination of the $m$th data point.
After being fitted the data
are re-plotted against $L/\xi$ to check that they
obey the scaling law (\ref{fssxi}).

We also need to determine the goodness of fit $Q$
and confidence intervals for the fitting parameters.
The goodness of fit measures the credibility of the fit;
$Q>0.001$ is often regarded as acceptable in other
applications \cite{NUMREP}.
We have checked that the numerical procedure used to estimate
the localization lengths does so with an error which is
approximately normally distributed.
If we ignore the presence of any systematic corrections to
scaling in the data, 
this permits the use of the $\chi^2$
likely-hood function to determine the ``best fit''
and the estimation of $Q$ from the $\chi^2$
distribution with $M-N$ degrees of freedom.

\begin{figure}
\epsfig{file=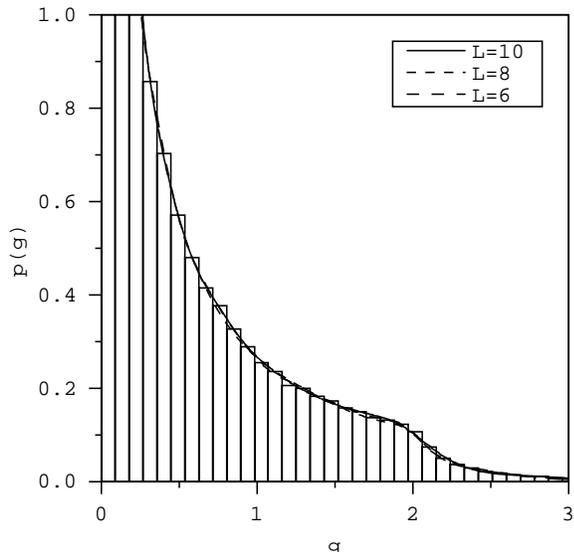,width=3.0in}
\caption{The critical conductance distributions for parameter
set Oa listed in Table ~\ref{T1}.}
\label{F2}
\end{figure}

The confidence intervals for the
fitted parameters were estimated in two independent ways:
first from the Hessian matrix obtained
in the least squares fitting procedure and
second using the Bootstrap procedure described
in \cite{NUMREP}.
In the latter method the original data are repeatedly randomly sampled 
(with replacement) and fitted.
This provides an independent check of the distribution
of the fitted parameters.
Both methods gave approximately the same results.
We chose to present the errors as $95.4\%$
marginal confidence intervals as given by the
Boostrap method.

The results are summarized in Tables \ref{T1}
and \ref{T2}.
Given the confidence intervals the probability
that the values of the
critical exponent for the orthogonal and unitary
universality classes are the same seems to be negligible.
Different values of $\Lambda_c$ can also be distinguished
confirming what is clearly evident in Fig. \ref{F1}
that the orthogonal and unitary data scale differently.
We conclude that the scaling function is sensitive to
the breaking of time reversal symmetry.

We now turn to the conductance distribution.
We consider a cubic ``sample'' of side $L$ attached
to semi- infinite leads on two opposite faces.
The 
disorder $W$ is set to zero in the leads,
the Fermi energy and magnetic field
are constant throughout.
The zero temperature linear conductance $G=(e^2/h)g$
can be obtained from
$g={\rm tr}tt^{\dagger}$
where $t$ is the transmission matrix  of the sample.
The $t$ matrix can be related to a Green's
function \cite{DATTA,ANDO}
which can be determined iteratively.

\begin{figure}
\epsfig{file=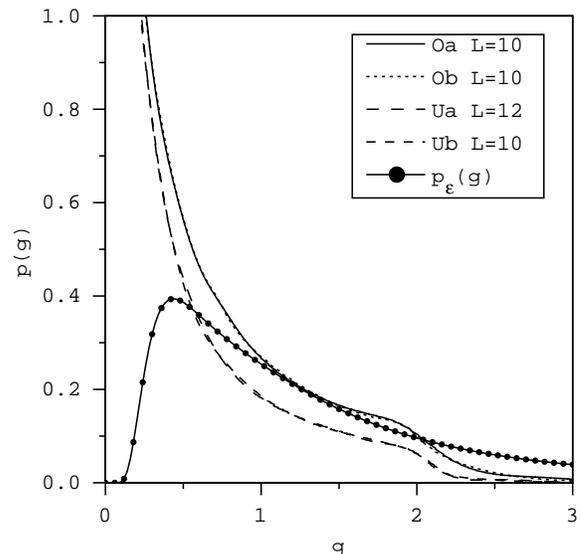,width=3.0in}
\caption{The distribution of $g$ at the critical point.
Orthogonal and unitary distributions can be clearly distinguished.
The critical conductance distribution obtained in
the $\epsilon$ expansion is also shown.}
\label{F3}
\end{figure}

For each set of parameters we have 
calculated the conductances for an ensemble of
$100,000$ realizations of the random potential.
Some typical results are presented in Fig. \ref{F2}.
Our analysis of these results
is based on the generalization of (\ref{fssxi})
$p(g) = p(g,L/\xi)$.
At the critical point $\xi$ diverges and we should obtain a
universal critical conductance distribution $p_c(g)$ which is
independent of  $L$.
The scale invariance of $p_c(g)$
is  clearly demonstrated, at least for the range of
system sizes studied,  in Fig. \ref{F2}.
We found a similar scale invariance for all the cases listed in
Table \ref{T1}.
In Fig. \ref{F3} we have plotted the critical conductance distributions
obtained  and
in Table \ref{T3} tabulated some averages of
these distributions.
The results are consistent with the existence
of distinct orthogonal and unitary critical conductance
distributions.

We now discuss the general features of $p_c(g)$
focusing on the orthogonal universality class and
making a comparison with the
critical distribution obtained in the $\epsilon$ expansion
in the field theory.
The $n$th cumulant $c_n$ of $p_c(g)$
for a $d$ dimensional cube of linear dimension $L$ is
\cite{AKL}
\begin{equation}
\begin{array}{ll}
c_n = \left( \frac{\pi}{2} \right)^{n} &
\left\{ \begin{array}{ll}
\epsilon^{n-2} & n \alt n_0 \\
(L/l)^{\epsilon n^2-2n} & n \agt n_0
\end{array}
\right.
\end{array}
\label{cumulants}
\end{equation}
where $\epsilon=d-2$,
$n_0$ is an integer of order $1/\epsilon$
and $l$ is the elastic mean free path.
As described in \cite{SHAPIRO} it is possible, under certain
assumptions, to derive $p_c(g)$ from this.
Extrapolating to $d=3$ we find
\begin{equation}
p_{\epsilon}(g) = \frac{1}{{\rm e}} \delta(g) +
\frac{\pi}{2\text{e}} \left(
u(\frac{\pi g}{2}) + \frac{1}{2!} [u*u](\frac{\pi g}{2}) + \dots
\right) ,
\label{series}
\end{equation}
where
$u(x) = 4 x^{-3}\exp(-2/x)$,
$*$ denotes the convolution and e=2.71828...
The series (\ref{series}) is easily
handled numerically.
The result is shown in Fig. \ref{F3}.

The most obvious feature is that $p_c(g)$
is {\it not} peaked about
its mean value $<g>$.
The conductance fluctuations, as measured by the standard
deviation $\sigma_g$, are of the same
order of magnitude as  $<g>$.
If we compare with $p_{\epsilon}(g)$
we find that we have a good approximation to
the central region of the distribution function
but a rather bad approximation to its tails.
The large $g$ tail of $p_{\epsilon}(g)$
decays as $1/g^3$ which means that all cumulants
higher than $c_1$ diverge.
This is reflected in (\ref{cumulants})
where these cumulants are not
universal but depend on $l$ and $L$.
We could find no evidence of this behavior, however, in the 
simulation; as seen in Fig. $\ref{F3}$ there is sharp
decay of $p_c(g)$ above $g\simeq 2$ and the higher cumulants,
at least as far as $n=4$, seem to have universal
values.

\begin{figure}
\epsfig{file=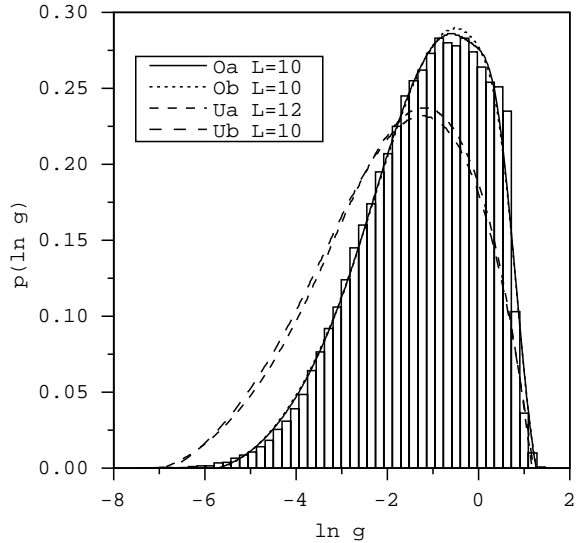,width=3.0in}
\caption{The critical distribution after changing variables.}
\label{F4}
\end{figure}

Another way to look at the critical distribution is to
change variables to $\ln  g$ (see Fig. \ref{F4}).
While the distribution is certainly not of Gaussian form,
it does show a central tendency.

In conclusion we have presented numerical evidence
which, we think, confirms that the critical behavior 
at the Anderson transition is sensitive to
perturbations which break time reversal symmetry.

T.O. would like to thank Yoshiyuki Ono for important discussions
at the outset of the present work.
Some of this work was carried out on supercomputer facilities at
the ISSP, Univ. Tokyo and
the Institute of Physical and Chemical Research.


\begin{table}
\begin{tabular}{|l|l|l|l|l|l|l|l|l|}
Label & $E_f$ & $\phi$  & $L$ & $[W_{min},W_{max}]$ & $M$ & $N$ &
$\chi^2_{\text{min}}$ & Q \\
\hline
Oa    & $0$   & $0$      & $\{6,8,10,12\}$     & $[15,18]$     & $86$ & $6$
& $79$ & 0.50\\
Ob    & $0.5$ & $0$      & $\{6,8,10\}$     & $[15.6,17.6]$ & $73$ & $6$ &
$49$ & 0.96 \\
Ua    & $0$   & $1/3$  &  $\{6,9,12\}$     & $[17,20]$     & $55$ &
$6$ & $35$ & 0.94\\
Ub    & $0$   & $1/4$  &  $\{8,10,12,14\}$ & $[17.5,19.5]$ & $66$ &
$6$ & $65$ & 0.31\\
\end{tabular}
\caption{The parameters used in the numerical study:
Fermi energy $E_f$, flux $\phi$, system sizes $L$,
the range of disorder, the number of data points $M$, 
the number of fitting parameters $N$, the value of $\chi^2$
for the best fit $\chi^2_{\text{min}}$ and the goodness of fit Q
as estimated from the $\chi^2$ distribution.}
\label{T1}
\end{table}

\begin{table}
\begin{tabular}{|l|l|l|l|}
Label & $W_c$ & $\Lambda_c$ & $\nu$ \\
\hline
Oa    & $16.448\pm.014$ & $0.5857\pm .0012$ & $1.59\pm .03$  \\
Ob    & $16.442\pm.018$ & $0.5862\pm .0015$ & $1.60\pm .06$  \\
Ua    & $18.316\pm.015$ & $0.5683\pm .0013$ & $1.43\pm .04$  \\
Ub    & $18.375\pm.017$ & $0.5662\pm .0016$ & $1.43\pm .06$  \\
\end{tabular}
\caption{The best fit values of the critical parameters together with their
$95.4\%$ confidence intervals as estimated using the bootstrap method.}
\label{T2}
\end{table}

\begin{table}
\begin{tabular}{|l|l|l|l|l|}
Label & $<g>$ & $\sigma_g$ & $<\ln  g>$ & $\sigma_{\ln  g}$ \\
\hline
Oa $L=10$  & $0.579$ & $0.598$  & $-1.206$ & $1.319$ \\
Ob $L=10$  & $0.578$ & $0.598$  & $-1.207$ & $1.318$ \\
Ua $L=12$  & $0.395$ & $0.499$  & $-1.869$ & $1.590$ \\
Ub $L=10$  & $0.395$ & $0.499$  & $-1.864$ & $1.583$ \\
\end{tabular}
\caption{The means and standard deviations of the critical
distribution of $g$ and $\ln  g$.}
\label{T3}
\end{table}


\begin{references}
\bibitem{KM} For a review see B. Kramer and A. MacKinnon,
Rep. Prog. Phys. {\bf 56}, 1469 (1993).
\bibitem{EFETOV} K. B. Efetov, Adv. in Phys. {\bf 32}, 53 (1983).
\bibitem{WEGNER} W. Benreuther and F. Wegner, Phys. Rev. Lett.
{\bf 57}, 1385 (1986).
\bibitem{HIKAMI} S. Hikami, Prog. Theo. Phys. Supp.
{\bf 107}, 213 (1992).
\bibitem{HKO} M. Henneke, B. Kramer and T. Ohtsuki, Europhys. Lett.
{\bf 27}, 389 (1994).
\bibitem{MAC} A. MacKinnon, J. Phys.: Condens. Matt. {\bf 6}, 2511 (1994).
\bibitem{OOK} T. Ohtsuki, B. Kramer and Y. Ono, J. Phys. Soc. Jpn. {\bf 62},
224 (1993).
\bibitem{CD} J.T. Chalker and A. Dohmen, Phys. Rev. Lett. {\bf 75},
 4496 (1995).
\bibitem{TOHRU} T. Kawarabayashi, T. Ohtsuki, K. Slevin
and Y. Ono, Phys. Rev. Lett. {\bf 77}, 3593 (1996).
\bibitem{HOFSTETTER} E. Hofstetter, cond-mat/9611060.
\bibitem{GOLD} Ch. 9, {\it Lectures on Phase Transitions and the Renormalization
Group}, N. Goldenfeld (Addison Wesley, 1992).
\bibitem{MK} A. MacKinnon and B. Kramer, Z. Phys. B {\bf 53},
1 (1983).
\bibitem{Imry} I. Lerner and Y. Imry, Europhys. Lett.
{\bf 29}, 49 (1995).
\bibitem{WEGNER2} F. Wegner, Z. Phys. B {\bf 25}, 327 (1976).
\bibitem{ZINN}  Ch. 28, {\it Quantum Field Theory and Critical Phenomena}
J. Zinn-Justin (Oxford, 1996).
\bibitem{NUMREP} Ch. 15, {\it Numerical Recipes in Fortran},
W. Press, B. Flannery and S. Teukolsky, (Cambridge Univ. Press, 1992).
\bibitem{DATTA} Ch. 3, {\it Electronic Transport in Mesoscopic Systems}
S. Datta (Cambridge Univ. Press, 1995).
\bibitem{ANDO} T. Ando, Phys. Rev. B {\bf 44}, 8017 (1991).
\bibitem{AKL} B.L. Altshuler, V. Kravtsov and I. Lerner,
Sov. Phys. JETP {\bf 64}, 1352
(1986);
Phys. Lett. {\bf A134}, 488 (1989).
\bibitem{SHAPIRO} B. Shapiro, Phys. Rev. Lett. {\bf 65}, 1510 (1990);
A. Cohen and B. Shapiro, Int. J. Mod. Phys. B{\bf 6}, 1243 (1992)
\end{references}
\end{document}